\documentclass[Times,twocolumn,tighten,deluxetable]{aastex63}

\usepackage{graphicx}
\usepackage{rotating}
\usepackage{multirow}

\submitjournal{ApJ}

\shorttitle{Accretion Flow Dynamics of EXO 1846-031}
\shortauthors{Nath et al.}

\begin{document}

\title{Accretion Flow Properties of EXO 1846-031 During its Multi-Peaked Outburst After Long Quiescence}

\correspondingauthor{Dipak Debnath}
\email{dipakcsp@gmail.com}
\email{sujoynath0007@gmail.com}

\author[0000-0002-6640-0301]{Sujoy Kumar Nath}
\affiliation{Indian Center for Space Physics,  466 Barakhola, Netai Nagar, Kolkata 700099, India}

\author[0000-0003-1856-5504]{Dipak Debnath}
\affiliation{Institute of Astronomy Space and Earth Science, AJ 316, Sector II, Salt Lake, Kolkata 700091, India}
\affiliation{Institute of Astronomy, National Tsing Hua University, Hsinchu 300044, Taiwan}

\author[0000-0002-6252-3750]{Kaushik Chatterjee}
\affiliation{South Western Institute for Astronomical Research, Yunnan University, University Town, Chenggong, Kunming 650500, P. R. China}
\affiliation{Institute of Astronomy Space and Earth Science, AJ 316, Sector II, Salt Lake, Kolkata 700091, India}
\affiliation{Institute of Astronomy, National Tsing Hua University, Hsinchu 300044, Taiwan}

\author[0000-0002-7658-0350]{Riya Bhowmick}
\affiliation{Indian Center for Space Physics,  466 Barakhola, Netai Nagar, Kolkata 700099, India}

\author[0000-0002-5617-3117]{Hsiang-Kuang Chang}
\affiliation{Institute of Astronomy, National Tsing Hua University, Hsinchu 300044, Taiwan}
\affiliation{Department of Physics, National Tsing Hua University, Hsinchu 300044, Taiwan}

\author[0000-0002-0193-1136]{Sandip K. Chakrabarti}
\affiliation{Indian Center for Space Physics,  466 Barakhola, Netai Nagar, Kolkata 700099, India}


\begin{abstract}
We study the recent outburst of the black hole candidate EXO 1846-031 which went into an outburst in 2019 
after almost 34 years in quiescence. We use archival data from Swift/XRT, MAXI/GSC, NICER/XTI and NuSTAR/FPM 
satellites/instruments to study the evolution of the spectral and temporal properties of the source during 
the outburst. Low energy (2-10 keV) X-ray flux of the outburst shows multiple peaks making it a multipeak outburst. 
Evolving type-C quasi-periodic oscillations (QPOs) are observed in the NICER data in the hard, hard intermediate 
and soft intermediate states. We use the physical Two Component Advective Flow (TCAF) model to analyze the combined 
spectra of multiple satellite instruments. According to the TCAF model, the accreting matter is divided into Keplerian 
and sub-Keplerian parts, and the variation in the observed spectra in different spectral states arises out of the variable 
contributions of these two types of accreting matter in the total accretion rate. Studying the evolution of the accretion 
rates and other properties of the accretion flow obtained from the spectral analysis, we show how the multiple peaks in the outburst 
flux arises out of variable supply of accreting matter from the pile-up radius. 
We determine the probable mass of the black hole to be $10.4^{+0.1}_{-0.2}~M_\odot$ from the spectral analysis with the TCAF model. 
We also estimate viscous time scale of the source in this outburst to be $\sim 8$ days from the peak difference of the Keplerian and 
sub-Keplerian mass accretion rates.

\end{abstract}


\section{Introduction}

A Low mass Black hole X-ray binary system (BHXRBs) consists of a stellar-mass main-sequence star orbiting around a 
stellar-mass black hole (SMBH). Transient BHXRBs spend most of their lifetime in a quiescent state, exhibiting very low X-ray 
luminosity ($L_X \sim 10^{30-33}$ ergs/s; Tetarenko et al. 2016). Occasionally transient BHXRBs show bright outbursts, 
lasting for a few weeks to a few months, during which the source becomes extremely luminous
($L_X \sim 10^{37-38}$ ergs/s; Tanaka \& Shibazaki 1996).

Due to its non-zero angular momentum, matter from the companion star accretes onto the black hole (BH), forming an inward-spiralling accretion disk.
The accumulating matter heats up the disk, and the matter in the disk gets ionized causing thermal-viscous instability (Dubus et al. 2001; Lasota 2001). 
As a result of the instability, the viscosity of the ionized matter in the outer disk increases suddenly. This causes more angular momentum 
to be redistributed outward, and the accretion rate in the inner disk increases rapidly, triggering an outburst 
(Chakrabarti \& Titarchuk 1995; Ebisawa et al. 1996; Chakrabarti 2013). During an outburst, low mass BHXRBs go through 
a succession of `accretion states', showing a rapid change in their temporal and spectral properties (Fender et al. 2004;
Homan \& Belloni 2005; McClintock \& Remillard, 2006). During the initial phase of the outburst, the source luminosity is 
low and the energy spectrum can be approximated with a hard non-thermal power-law component. This state is called 
the hard state (HS). As the outburst progresses, the source transits through the hard-intermediate state (HIMS) and 
soft-intermediate state (SIMS), when the source luminosity gradually increases and the contribution of the low energy thermal
photons increase, which gradually softens the spectrum. The source luminosity becomes maximum in the soft state (SS) when the spectrum is dominated by 
a thermal multicolor disk blackbody. After that, the source luminosity gradually decreases, and the source transits through 
SIMS, HIMS and finally, to the HS. Low-frequency peaked and narrow noise components called quasi-periodic oscillations (QPOs) 
has been observed in the power-density spectra (PDS) of most BHXRBs. Their properties (centroid frequency, Q-value, rms amplitude 
and noise) also vary depending on the spectral state, and Casella et al. (2005) have classified these LFQPOs into three types:
A, B, and C. Generally, type-C QPOs with monotonically increasing or decreasing centroid frequency can be observed in the HS and
HIMS, while no QPOs are observed in the SS. The evolution of these spectral and temporal properties are strongly correlated, which is 
manifested in the `Hardness-Intensity Diagram' (HID; Belloni et al. 2005; Debnath et al. 2008) or the `Accretion Rate Ratio-Intensity Diagram' 
(ARRID; Jana et al. 2016). 

Two separate mechanisms are responsible for the production of low and high-energy X-ray radiation from the accretion disks.
An optically thick, geometrically thin Keplerian flow dissipates the gravitational energy of the accreting matter through 
viscosity and emits multicolor thermal blackbody photons (Novikov \& Thorne 1973; Shakura \& Sunyaev 1973). 
When these low-energy photons get intercepted by a hot electron cloud, they get repeatedly inverse Comptonised and are 
emitted as high-energy X-rays (Sunyaev \& Titarchuk 1980, 1985). While there is general agreement about the emission mechanisms,
the actual nature of the hot electron cloud or the Compton cloud has been a matter of debate. 
According to the Two-Component Advective Flow (TCAF) model (Chakrabarti \& Titarchuk 1995; Chakrabarti 1997, Chakrabarti 2018),
the CENtrifugal pressure supported BOundary Layer (CENBOL) acts as the Compton cloud. This CENBOL is formed near the black hole when the
low viscous, freely falling sub-Keplerian matter comes to a halt as the outward centrifugal pressure becomes comparable to the inward 
gravitational force, and it forms a standing or oscillating shock. The post-shock region becomes hot and puffs up and forms a torus-like region of 
hot ionised matter. In the equatorial region, the viscosity remains higher than a certain critical value to maintain Keplerian angular momentum, 
and this Keplerian matter becomes optically thick and emits the multicolor soft photons which are then partially intercepted by the CENBOL 
and emitted as hard non-thermal photons. In the TCAF model, any observed spectrum depends on four independent flow parameters, 
i.e. the accretion rates of the Keplerian and the sub-Keplerian matter, the location of the shock which is the outer boundary of CENBOL, 
and the ratio of the post-shock to pre-shock matter densities (compression ratio). Additionally, it also depends on the mass of the BH 
and a normalization factor which is the ratio of emitted to observed X-ray spectrum, both of which are constants for a given source.
As an outburst starts, the faster and hotter sub-Keplerian matter rushes towards the BH and forms the CENBOL which increases the hard Comptonised
flux. The Keplerian matter, which has a low velocity due to the higher viscosity, gradually moves towards the BH and cools down the CENBOL.
The CENBOL region shrinks in size, the hard photon flux decreases and the spectra become gradually softer. As the outer boundary of the CENBOL 
oscillates (e.g. due to a resonance between the Compton cooling and compressional heating), the Comptonized hard X-ray intensity also varies
which gives rise to the observed quasi-periodic oscillations. This CENBOL also acts as the base of the jet/outflows.

To study how the physical flow parameters vary during an outburst and to estimate the intrinsic parameters of the BH, this TCAF model has been 
incorporated into the spectral analysis software package {\fontfamily{pcr}\selectfont XSPEC} (Arnaud, 1996) as a local additive table model 
(Debnath et al. 2014, 2015). So far, the TCAF model has been used to study the accretion flow dynamics of more than fifteen BHXRBs 
(Mondal et al. 2016; Debnath et al. 2017; Chatterjee et al. 2021). Intrinsic parameters, like the 
BH mass and its distance have been estimated (Molla et al. 2017; Chatterjee et al. 2019; Jana et al. 2020a; Nath et al. 2023). The origin of QPOs 
and jet/outflows has also been successfully studied using this model (Mondal et al. 2015; Chakrabarti et al. 2015; Chatterjee et al. 2016, Jana et al. 2017; Debnath et al. 2021)

Galactic X-ray transient EXO 1846-031 was first discovered by EXOSAT during its outburst in 1985 (Parmar \& White 1985).
Based on the ultra-soft component in the spectra of this outburst, Parmar et al. (1993) indicated the source EXO 1846-031 is a BH candidate.
After the first outburst, the source remained in quiescence for almost 34 years. Recently, the source was again found to be in outburst by 
MAXI on 2019 July 23 (Negoro et al. 2019). Evolving Type-C QPOs were observed in the Insight-HXMT and NICER data 
(Liu et al. 2021) which is generally observed in BHXRBs. From strong reflection features in the NuSTAR spectrum, Draghis et al. (2020) 
suggested EXO 1846-031 to be a BH with nearly maximal spin ($a=0.99^{+0.002}_{-0.001}$) with a disk inclination of $\theta\approx73^{\circ}$ and a mass of $9\pm5 ~M_\odot$. 
From Insight-HXMT and NuSTAR data, Wang et al. (2021) found signatures of an ionised disk wind with velocities up to $0.06c$. 
They suggest EXO 1846-031 is a low inclination system with $\theta\approx40^{\circ}$. Ren et al. (2022) investigated the spectral evolution from 
Insight-HXMT data and suggested that the maximal spin found by Draghis et al. (2020) might be affected by choice of a different 
hardening factor ($f_{col}$). Evidence of the presence of a pair of 3:2 ratio high-frequency quasi-periodic oscillations (HFQPO)
was found, and based on this the probable mass of the source was determined to be $3.4\pm0.2 ~M_\odot$ (Strohmayer \& Nicer Observatory
Science Working Group 2020).
Analysing the radio observations from MeerKAT, VLA and AMI-LA, Williams et al. (2022) suggested a distance range of 
2.4–7.5 kpc, and a jet speed of $\beta_{int}=0.29c$.

We study the spectral and temporal properties of EXO 1846-031 during its 2019 outburst using Swift/XRT, Swift/BAT, 
MAXI/GSC, NICER/XTI and NuSTAR/FPM data with the TCAF model in this paper. We discuss the observation, data reduction, 
and analysis procedures briefly In \S2. In \S3 we present the results of our analysis. In \S4, we discuss the 
obtained results and draw conclusions.

\vskip 1.0cm

\section{Observation and Data Analysis}

\subsection{Observations}

We study the $2019-2020$ outburst of EXO 1846-031 using archival data from Swift (Gehrels et al. 2004), NICER (Gendreau et al. 2012), 
MAXI (Matsuoka et al. 2009), and NuSTAR (Harrison et al. 2013) satellites. We study the evolution of the X-ray fluxes in the soft
and hard energy bands and their ratios using MAXI/GSC ($2-10$~keV) and Swift/BAT ($15-50$~keV) data of $\sim 10$~months from 2019 July 9 (MJD=58673) 
to 2020 April 10 (MJD=58949). For the detailed temporal and spectral study, we use data from Swift/XRT, NICER/XTI, MAXI/GSC and NuSTAR/FPM
satellites/instruments.

Although NICER and Swift monitored the source regularly in the rising phase of the outburst, during the declining phase, there is a data gap
of $\sim 3$~months for Swift and $\sim 4$~months for NICER. We use 14 data of NICER/XTI ($1-11$ keV) and 11 data of Swift/XRT ($1-10$ keV) 
for spectral analysis. To study the spectra in a wider energy band, we also use MAXI/GSC ($7-20$ keV) and NuSTAR/FPM ($4-79$ keV) simultaneously 
with NICER and Swift data. A detailed log of the observations is given in Table 1.

\subsection{Data Reduction}

\subsubsection{Swift}
Swift/XRT window timing (WT) mode data were used in our analysis. Level 1 data files obtained from the archive are processed with the {\fontfamily{qcr}\selectfont XRTPIPELINE} task to
produce Level 2 clean event files. A circular region of radius $30''$ around the source location is then used to extract the source
spectra and a region of the same radius is chosen away from the source to extract the background spectra using the tool {\fontfamily{qcr}\selectfont XSELECT}. ARF files
are created using the tool {\fontfamily{qcr}\selectfont XRTMKARF} and corresponding RMFs are obtained from the {\fontfamily{qcr}\selectfont CALDB}.
Using the {\fontfamily{qcr}\selectfont GRPPHA} tool, the spectra are rebinned to have at least 20 counts/bin.
Swift/BAT daily lightcurves are obtained from the Swift  \href{https://swift.gsfc.nasa.gov/results/transients/weak/EXO1846-031/}{website}.

\subsubsection{NICER}
NICER is an external payload attached to the International Space Station which has an X-ray timing instrument (XTI; Gendreau et al. 2012) 
working in the energy range 0.2-12 keV with a timing resolution of $\sim100$ ns and spectral resolution of $\sim85$ eV at 1 keV. For analysis,
the Level 1 data files are processed with {\fontfamily{qcr}\selectfont nicerl2} script in the latest caldb environment (ver. xti20221001) to obtain
Level 2 clean event files. The command {\fontfamily{qcr}\selectfont barycorr} is then used to apply barycentric correction to the event files.
The lightcurves and spectra are extracted from these barycentre-corrected event files using the {\fontfamily{qcr}\selectfont XSELECT} task. 
The {\fontfamily{qcr}\selectfont nibackgen3C50} tool (Remillard et al. 2022) is then used to simulate the background corresponding to each observation.
The spectra are then rebinned to have at least 20 counts/bin with the {\fontfamily{qcr}\selectfont GRPPHA} task.

\subsubsection{NuSTAR}
NuSTAR raw data from the web archive is reduced with the NuSTAR data analysis software ({\fontfamily{qcr}\selectfont NuSTARDAS}, version 1.4.1).
Cleaned event files are produced using the {\fontfamily{qcr}\selectfont nupipeline} task in the presence of the latest calibration files.
With the {\fontfamily{qcr}\selectfont XSELECT} task, a circular region of $60''$ centred at the source coordinates is chosen as the source region, 
and a circular region with the same radius away from the source location is chosen as the background region.
The {\fontfamily{qcr}\selectfont nuproduct} task is then used to extract the spectrum, ARF and RMF files. 
The extracted spectra are then rebinned to have at least 30 counts/bin with the {\fontfamily{qcr}\selectfont GRPPHA} task.

\subsubsection{MAXI}
MAXI/GSC spectra are obtained using the \href{http://maxi.riken.jp/mxondem/}{MAXI on-demand} process web tool (Matsuoka et al. 2009).
To study the evolution of the X-ray fluxes, daily average lightcurves are obtained from the MAXI \href{http://maxi.riken.jp/star_data/J1849-030/J1849-030.html}{website}.

\begin{figure*}
\vskip 0.5cm
  \centering
    \includegraphics[angle=0,width=14cm,keepaspectratio=true]{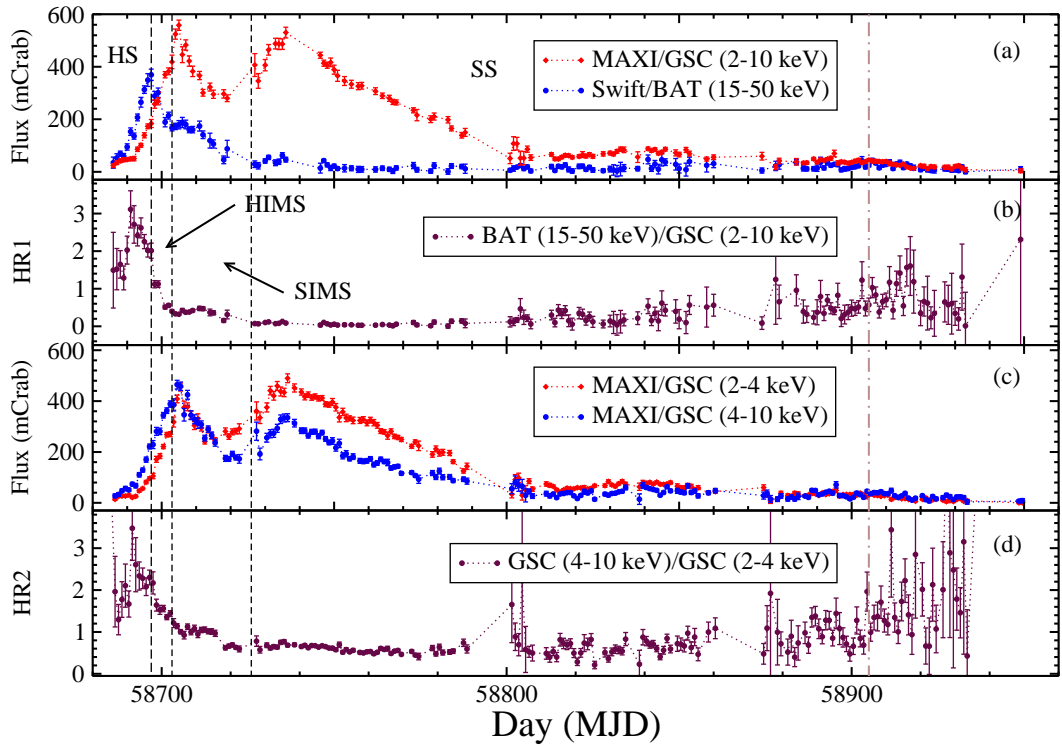}
	\caption{Variations of (a) 2-10 keV MAXI/GSC flux and 15-50 keV Swift/BAT flux in units of mCrab 
	                       (b) HR1, i.e., the ratio of 15-50 keV Swift/BAT flux to 2-10 keV MAXI/GSC flux 
			       (c) MAXI/GSC flux in 2-4 keV and 4-10 keV range in units of mCrab  
			       (d) HR2, i.e., the ratio of 4-10 keV to 2-4 keV MAXI/GSC flux are shown with time (in MJD).
                    Vertical dashed (online black) lines mark the transition between spectral states. 
                    The vertical dot-dashed (online pink) line marks the probable start of the declining hard state (see text for details).}
\end{figure*}

\subsection{Data Analysis}
Daily average light curve data of MAXI/GSC and Swift/BAT are used to study the variation of the X-ray flux in various energy bands 
throughout the outburst. To study the hardness variations, we use two types of hardness ratios, namely hardness ratio 1 (HR1) 
i.e. the ratio of 15-50 keV Swift/BAT flux in mCrab to 2-10 keV MAXI/GSC flux in mCrab, and hardness ratio 2 (HR2) i.e. the ratio
of 4-10 keV to 2-4 keV MAXI/GSC flux. To search for the presence of LFQPOs, we use the {\fontfamily{qcr}\selectfont powspec} task 
to generate power density spectra (PDS) from 0.01 s time binned light curves of NICER. The light curve of a total observation is separated 
into a number of intervals, each of which contains 8192 newbins. For each interval, a PDS is created, and these individual PDSs 
are averaged to generate a single PDS which was then geometrically rebinned. We model the PDSs with multiple Lorentzian models
in {\fontfamily{qcr}\selectfont XSPEC} version 12.11.1 (Arnaud 1996) to account for the broadband noise, QPOs and its harmonics.
From the fits we obtain the QPO frequencies ($\nu_{QPO}$), width ($\Delta\nu$), Q-value ($Q=\nu_{QPO}/\Delta\nu$) and RMS ($\%$) amplitude.

We utilize HEASARC's spectral analysis software package {\fontfamily{qcr}\selectfont XSPEC} version 12.11.1 (Arnaud 1996)
for analyzing the spectra. The spectra are fitted with the TCAF model based local additive table model (Debnath et al. 2014).
To fit spectra using the TCAF model, four input flow parameters are essential:
(\romannumeral 1) the Keplerian disk accretion rate ($\dot{m}_d$ in $\dot{M}_{Edd}$), 
(\romannumeral 2) the sub-Keplerian halo accretion rate ($\dot{m}_h$ in $\dot{M}_{Edd}$), 
(\romannumeral 3) the shock location ($X_s$ in Schwarzschild radius $r_s=2 GM_{BH}/c^2$), and 
(\romannumeral 4) the dimensionless shock compression ratio ($R = \rho_+/\rho_-$, ratio of the post-shock to the pre-shock matter density).
In addition, one system parameter, i.e., the mass of the BH ($M_{BH}$ in $M_\odot$) and one instrument parameter, i.e. the model normalization ($N$)
are required. To account for the interstellar absorption, we use the {\fontfamily{qcr}\selectfont TBabs} model with {\fontfamily{qcr}\selectfont vern} 
cross-sections (Verner et al. 1996) and {\fontfamily{qcr}\selectfont wilm} abundances (Wilms et al. 2000).
We use the {\fontfamily{qcr}\selectfont smedge} model to account for the instrumental features in the NICER spectra at $\sim1.8$ keV.

\section{Results}
After almost 34 years in quiescence, EXO 1846-031 again went into an outburst on 2019 July 23 (MJD 58687) which lasted for $\sim10$ months.
To examine the nature of the outburst and the accretion flow properties during the outburst, we carried out a detailed temporal and 
spectral study using data from multiple satellites. The results of the study are presented below. 

\subsection{Temporal Properties}
To study the outburst profile in different energy bands and the variation of hardness ratios, we use MAXI/GSC and Swift/BAT
daily light curve data. To study the low timescale variability features, we use NICER/XTI data due to its high 
temporal resolution.

\begin{figure}[h]
\vskip -0.2cm
  \centering
    \includegraphics[angle=0,width=9cm,keepaspectratio=true]{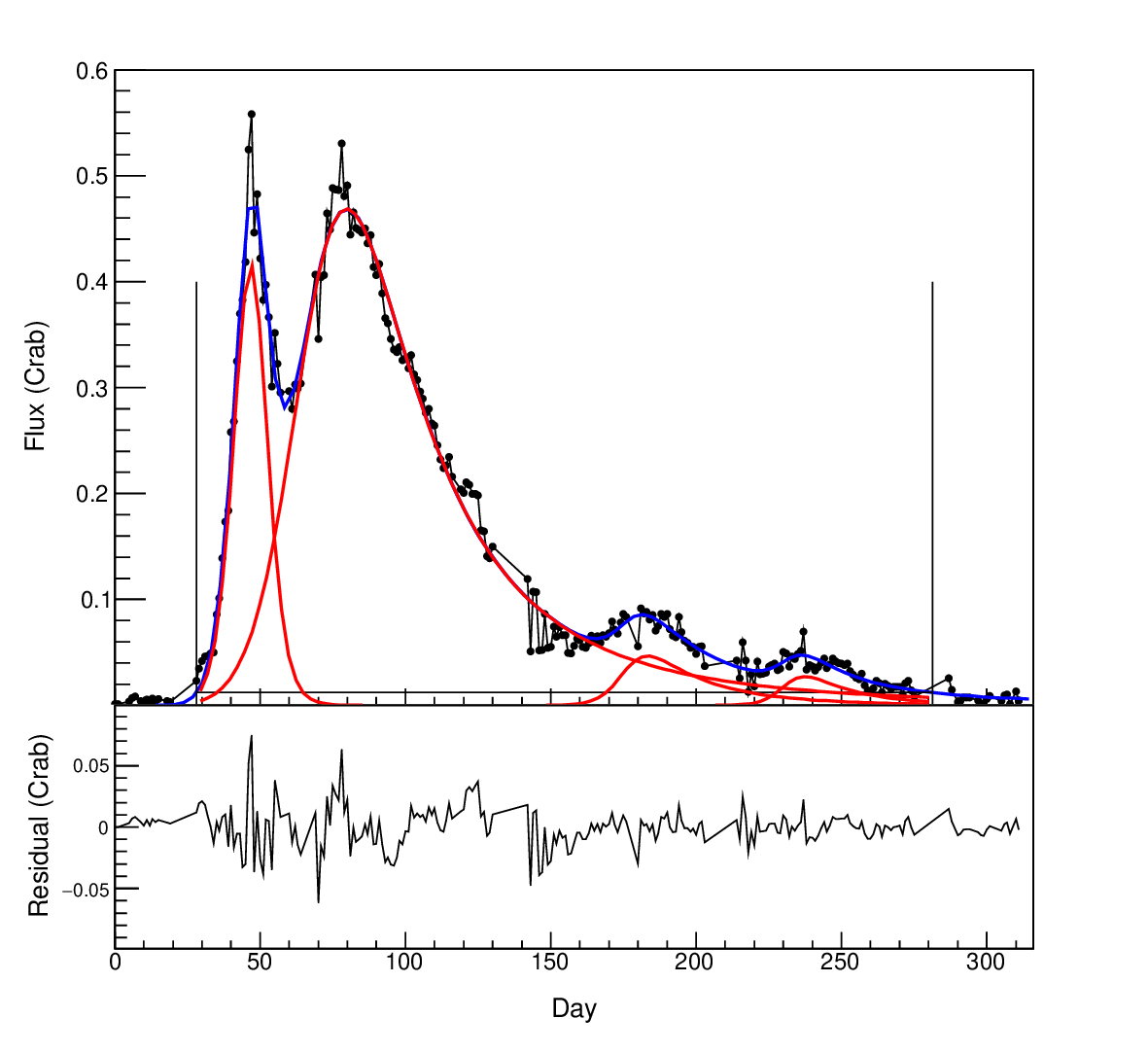}
	\caption{FRED profile fitted lightcurve of the outburst using four FRED profiles. The blue curve shows the combined FRED-fitted 
 curve and the four red curves show the FRED-fitted individual peaks. The flux value of quiescence ($12$ mCrab) is shown by a horizontal 
 black line, above which is considered as the outbursting phase. Two vertical black lines show the start and end of the outburst.
 Here MJD= 58658.5 is considered as the 0th day.}
\end{figure}

\subsubsection{Outburst Profile and Hardness Ratios}
We show the variation of X-ray fluxes in different energy bands and their hardness ratios 
from 2019 July 9 (MJD=58673) to 2020 April 10 (MJD=58949) in various panels of Fig. 1.
The variation of the Swift/BAT (15-50 keV) flux and the MAXI/GSC (2-10 keV) flux is shown in panel (a),
while the variation of their hardness ratio (HR1) is shown in panel (b). Likewise, panel (c) shows 
the variation of MAXI/GSC flux in lower (2-4 keV) and higher (4-10 keV) energy bands while panel (d) 
shows the variation in their hardness ratio (HR2). From the Figure, we can observe that at the start 
of the outburst, both soft and hard fluxes increased rapidly, and the 15-50 keV Swift/BAT flux
attained a maximum on MJD 58697, roughly 8 days before the softer (2-4 keV and 4-10 keV) MAXI/GSC fluxes.
The hardness ratios (HR1 and HR2) also increased and attained a maximum around MJD 58691 and decreased 
quickly to a low level. After the initial maximum, the Swift/BAT flux slowly decreased and decayed into 
quiescence at the end of the outburst. On the other hand, after the maximum around MJD 58705, the MAXI/GSC 
fluxes (in different energy bands) decreased for $\sim13$ days and then started to increase again. They 
attained a maximum around MJD 58736 and then decreased with an almost constant rate for $\sim65$ days. After 
that, the GSC fluxes remained at a constant low level till the end of the outburst.

Looking at the outburst profile, we can see that this 2019 outburst of EXO 1846-031 has shown two stronger flux peaks in the rising phase 
and two weaker peaks in the declining phase. To estimate the total amount of flux released during each of the peaks, we fit the $2-10$~keV 
MAXI/GSC lightcurve using FRED profiles (Kocevski et al. 2003). A combination of four FRED proﬁles are used to ﬁt the complete outburst (Fig. 2) 
(see, Chakrabarti et al. 2019, Bhowmick et al. 2021, Chatterjee et al. 2022 for more details). In the Fig. 2, the blue curve marks the combined 
fit of the entire outburst and the red curves mark individual FRED fitted peaks of the outburst. We choose $12$ mCrab as the threshold of flux 
for the outburst (Chakrabarti et al. 2019, Bhowmick et al. 2021, Chatterjee et al. 2022). The horizontal black line indicates the $12$ mCrab ﬂux value. Two vertical lines mark the start and the end of the outburst when 
the X-ray flux rises above and below this $12$ mCrab threshold. The total integrated X-ray ﬂux (IFX$_{tot}$) of the complete outburst calculated 
from the combined fit is $39.70^{+3.29}_{-3.05}$ Crab day. The individual integrated ﬂux values (IFX) of the first, second, third and fourth peaks are 
$6.31^{+0.26}_{-0.25}$ Crab day, $30.82^{+2.60}_{-2.42}$ Crab day, $1.77^{+0.60}_{-0.38}$ Crab day and $0.80^{+0.01}_{-0.16}$ Crab day respectively. IFX values depict 
the amount of energy release in each peaks. Comparing the IFX values of the four peaks, we can conclude that maximum amount of energy was released  
during the second peak, i.e., maximum amount of matter got cleared during the time period of the second peak of the outburst. 
 
\subsubsection{Power Density Spectra}
To study the presence and evolution of LFQPOs during the outburst, we use 0.01 s time-binned NICER 
light curves. We use zero-centred Lorentzian models to fit the broad noise components and narrow 
Lorentzians to fit the QPO peaks to determine the centroid frequencies, Q-values, rms amplitudes, etc.
We find the presence of QPOs in 19 NICER observations in the initial phase of the outburst. 
The observed QPOs can be classified as type-C which are characterized by a Q-value of $\sim 7-12$ and
an amplitude of $\sim3–16~\%$ rms that are superposed on a broad flat-top noise (Casella et al. 2005).
Figure 3 shows a representative PDS where a QPO of $3.24\pm0.03$ Hz can be seen along with its harmonic at $6.52\pm0.16$ Hz.
The QPOs are found in the hard, the hard-intermediate and the soft-intermediate states which are discussed in detail in later sections.

\begin{figure}[h]
    \includegraphics[angle=270,width=0.48\textwidth]{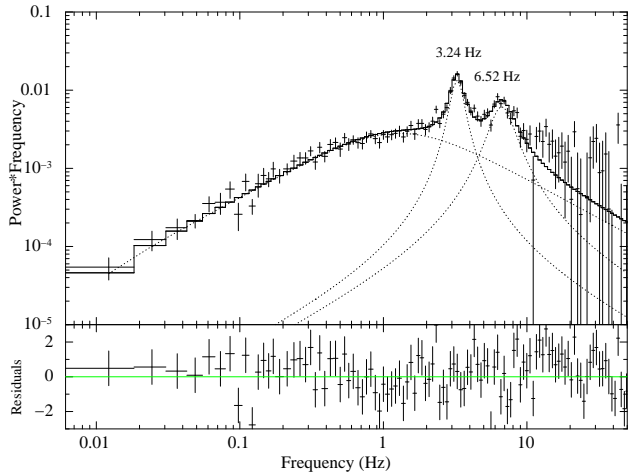}
	\caption{Continuum fitted PDS obtained from the NICER observation on 2019 August 05 and the best-fit Lorentzian components. 
	A QPO of frequency 3.24 Hz can be seen along with its harmonic at 6.52 Hz.}
\end{figure}

\subsection{Spectral Properties}
We use data from Swift/XRT, NICER/XTI, MAXI/GSC and NuSTAR/FPM for spectral analysis in a broad
1-79 keV energy range. We mainly use the absorbed TCAF model to study the spectra. We use the
{\fontfamily{qcr}\selectfont TBabs} model for absorption where the hydrogen column density ($N_H$)
was kept free. We found the $N_H$ to vary between $5.12\times10^{22}$ $cm^{-2}$ and
$10.94\times10^{22}$ $cm^{-2}$ during our analyses period. In the NICER spectra, edge-like
residuals are seen at $\sim1.8$ keV corresponding to the Silicon K edge which is an instrumental
feature typical for Si-based detectors (Alabarta et al. 2020, Miller et al. 2018).
We use the {\fontfamily{qcr}\selectfont smedge} model to account for it. An Fe-K$\alpha$ emission line at $\sim6.5$ keV
is also observed in the NICER spectra of the initial rising phase which was fitted using the 
{\fontfamily{qcr}\selectfont Gaussian} model. We jointly fit the XRT+GSC spectra with {\fontfamily{qcr}\selectfont constant*TBabs*(TCAF)}
model (Fig. 4a) and the NICER+GSC spectra with {\fontfamily{qcr}\selectfont constant*TBabs*smedge(TCAF)} or
{\fontfamily{qcr}\selectfont constant*TBabs*smedge(TCAF+Gaussian)} model (Fig. 4b). In the NICER+NuSTAR spectra,
a dip is observed at $\sim10$ keV in the NuSTAR data. At first, we fit the spectrum with {\fontfamily{qcr}\selectfont constant*TBabs*smedge(TCAF+Gaussian)} 
model ignoring the dip, and obtain $\chi^2/DOF=1.79$. To improve the statistic, we use the {\fontfamily{qcr}\selectfont gabs} model to account for the dip
and get a good statistic with $\chi^2/DOF=0.91$. The corresponding spectra are shown in Fig. 5(a--b). To investigate if this dip is a sign of disk 
reflection, we fit this spectrum with {\fontfamily{qcr}\selectfont constant*TBabs*smedge(diskbb+relxill)} and {\fontfamily{qcr}\selectfont constant*TBabs(diskbb+relxill)} models. The corresponding spectra are shown in Fig. 5(c--d). Detailed results of our spectral analysis are shown in Table 2.

\begin{figure*}
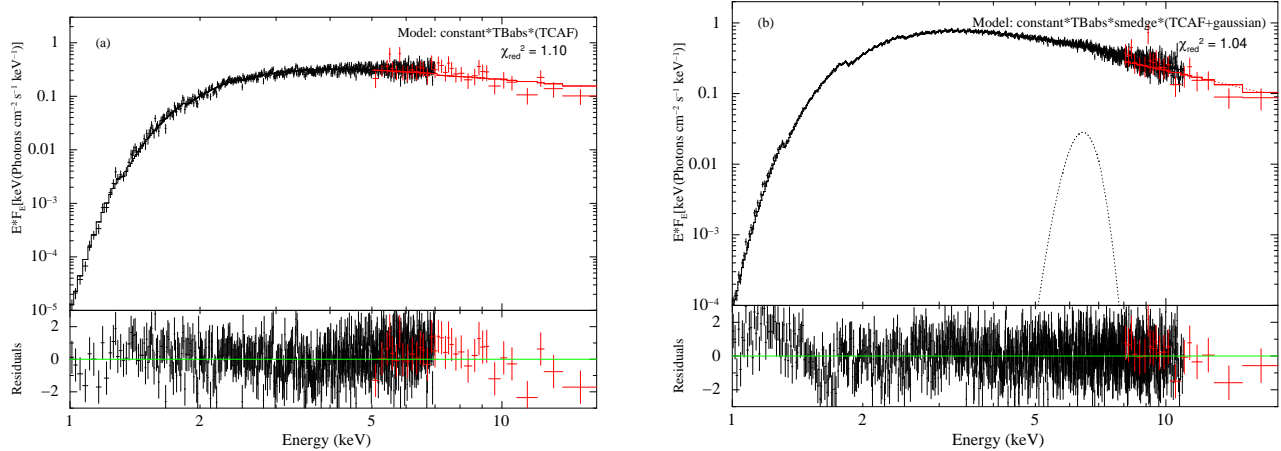

\vskip 0.2cm
\centering
\vbox{
\includegraphics[width=6.0truecm,angle=-90]{fig4a.eps}\hskip 0.5cm 
\includegraphics[width=6.0truecm,angle=-90]{fig4b.eps}}
\caption{Model fitted (a) XRT+GSC spectra (MJD 58697) (b) NICER+GSC spectra (MJD 58702) in $1-20$ keV energy band.}
\end{figure*}

\subsubsection{Evolution of the Spectral states}
The evolution of various temporal and spectral parameters of our analysis with the TCAF model shows that the source has evolved through 
different spectral states in this outburst. We get a rough estimation of the state evolution from the outburst profile and the variation of HRs. 
From the variation of the spectral parameters, we get a clearer picture of the state evolution as they show the actual evolution 
of the accretion flow dynamics, e.g. the change in the disk and halo accretion rates, the change of the position of the shock and its strength, etc. 
In the Figure 6, we show the variation of the disk accretion rate ($\dot{m}_d$), the halo accretion rate ($\dot{m}_h$), the total accretion rate 
($\dot{m}_d$ + $\dot{m}_h$) and the accretion rate ratio (ARR = $\dot{m}_h$/$\dot{m}_d$). In the Figure 7, we show the variation of the best fitted 
mass parameter ($M_{BH}$), the shock location ($X_s$), the shock compression ratio ($R$) alongwith the evolution of the QPO centroid frequency.
Here we discuss the spectral states in detail.

\textit{(\romannumeral 1) Rising Hard State (HS):}
The source was in the hard state when we start our spectral analysis on 2019 July 31 (MJD 58695). 
The total accretion rate was high in this state, and the maximum part of the accreting matter was sub-Keplerian
as the $\dot{m}_h$ was higher than the $\dot{m}_d$ by almost $\sim3$ times. The ARR was also high in this state, 
which started to decrease gradually as $\dot{m}_d$ started to increase and $\dot{m}_h$ started to decrease 
as the outburst progressed. The shock started to move towards the BH from a faraway location ($460r_s$), 
but its strength was almost constant in this period ($R\sim1.5$). Two LFQPOs were found in this state 
whose centroid frequency increased from $0.25$ Hz to $0.41$ Hz. High HR was also observed in this state 
as the hard flux (Swift/BAT) dominated the soft flux (MAXI/GSC). The source remained in this state until 2019 August 2 (MJD 58697).

\begin{figure*}[t]
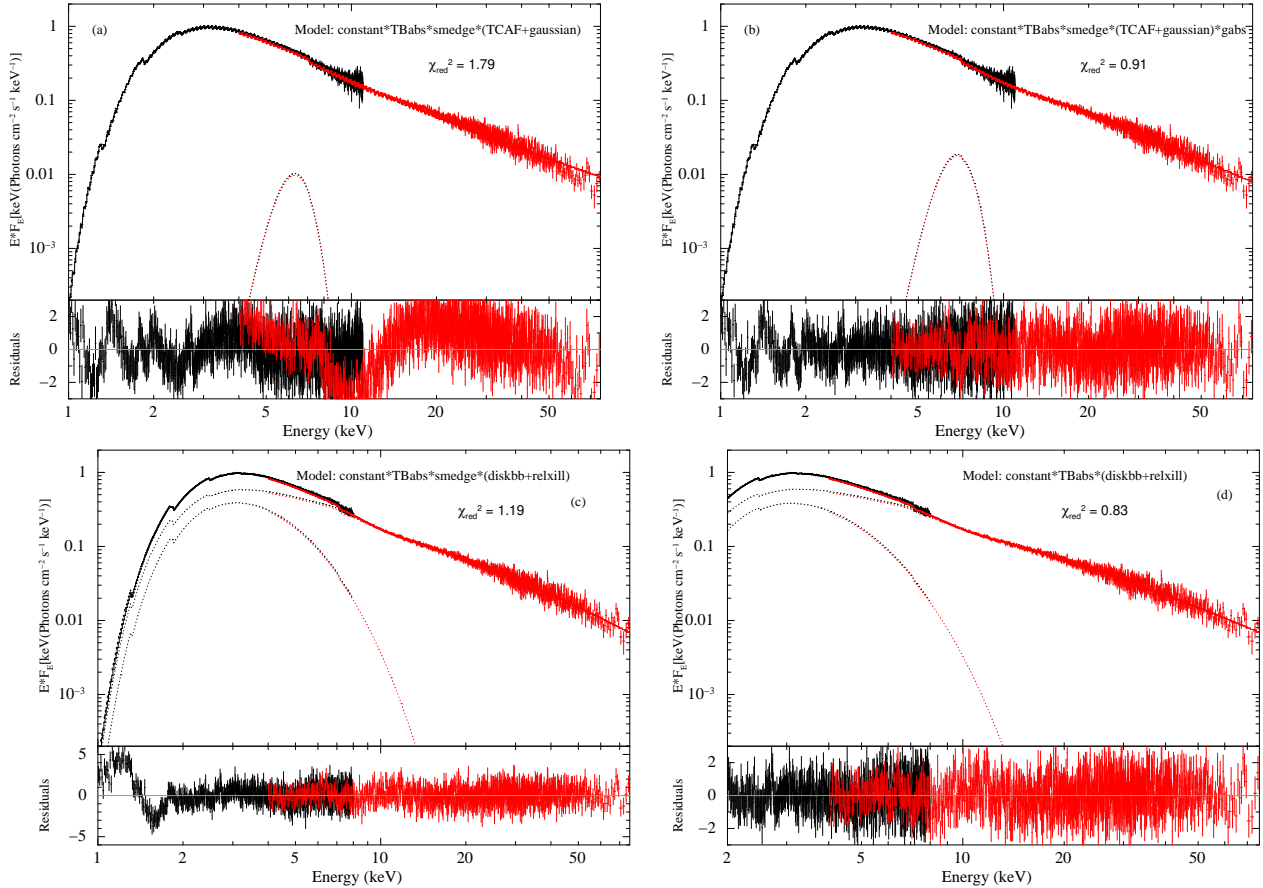

\vskip 0.4cm
\centering
\vbox{
\includegraphics[width=5.8truecm,angle=-90]{fig5a.eps}\hskip 0.1cm 
\includegraphics[width=5.8truecm,angle=-90]{fig5b.eps}}
\vskip 0.1cm
\vbox{
\includegraphics[width=5.8truecm,angle=-90]{fig5c.eps}\hskip 0.1cm 
\includegraphics[width=5.8truecm,angle=-90]{fig5d.eps}}
\caption{NICER+NuSTAR spectra on MJD 58708 fitted using different models.}
\end{figure*}

\textit{(\romannumeral 2) Rising Hard Intermediate State (HIMS):}
After MJD 58697, the total accretion rate started to decrease as the previously dominant $\dot{m}_h$ 
started to decrease rapidly. The total accretion rate began to increase again after 2019 August 5 (MJD 58700)
as $\dot{m}_d$ started to increase and became dominant. The ARR decreased steadily in this state. Likewise, 
the shock moved inward rapidly, moving from $\sim325~r_s$ to $\sim37~r_s$ in $\sim7$ days with decreasing strength.
Nine LFQPOs were found in this state whose centroid frequency increased rapidly to $\sim7$ Hz. 
The HR decreased in this state as the dominating hard flux began to decrease and soft flux increased steadily.
The source stayed in this state until 2019 August 8 (MJD 58703).

\textit{(\romannumeral 3) Rising Soft Intermediate State (SIMS):}
The total accretion rate decreased and became roughly constant at a low level after MJD 58703.
Both the $\dot{m}_d$ and $\dot{m}_h$ became steady, with $\dot{m}_d$ dominating over the $\dot{m}_h$.
The shock ceased to move towards the BH and came to a halt at $\sim35$$r_s$ and its strength also became constant.
We found eight LFQPOs during the initial part of this state, with their centroid frequency showing a slowly 
decreasing trend. The hard flux and the soft flux both decreased in this state, causing the HR to become low.
This state of the outburst continued until 2019, August 31 (MJD 58726).

\begin{figure*}[t]

  \centering
    \includegraphics[angle=0,width=14cm,keepaspectratio=true]{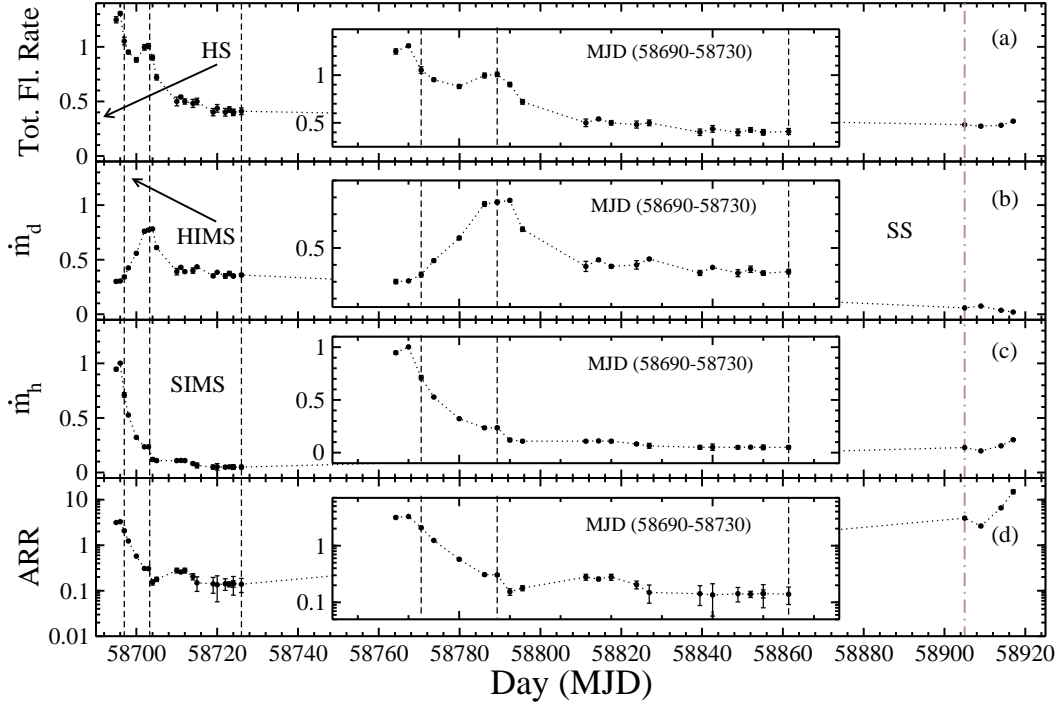}
	\caption{Variations of TCAF model fitted (a) total accretion rate ($\dot{m}_d$ + $\dot{m}_h$) in units of $\dot{M}_{Edd}$,
						 (b) Keplerian disk accretion rate ($\dot{m}_d$) in units of $\dot{M}_{Edd}$,/
						 (c) sub-Keplerian halo accretion rate ($\dot{m}_h$) in units of $\dot{M}_{Edd}$,
					     and (d) the accretion rate ratio (ARR = $\dot{m}_h$/$\dot{m}_d$)
	        are shown with time (MJD).
                           Vertical dashed (online black) lines mark the transition between spectral states. 
                    The vertical dot-dashed (online pink) line marks the probable start of the declining hard state (see text for details).}
\end{figure*}

\textit{(\romannumeral 4) Soft State/High Soft State (SS/HSS):}
After MJD 58726, the soft fluxes began to increase rapidly again which is quite unusual. An abrupt change has 
taken place in the accretion process. The hard 15-50 keV flux remained low, and this shows that the change 
in the accretion process has only affected the fluxes below 10 keV. The soft fluxes increased up to 
2019 September 10 (MJD 58736) and then decreased almost linearly until 2019 November 14 (MJD 58801) 
and became steady at a low level. The HRs also became low, signifying the source had transitioned into a
soft state/high soft state. Although XRT and NICER spectra were available for some days at the start of this
state, the TCAF fit of these spectra was statistically unacceptable, which shows that the two component configuration
of the accretion flow had been violated. We discuss this in detail in \S4. After November 2019, spectral data is unavailable
for $\sim$ 4 months, due to the source becoming sun-constrained (Williams et al. 2022). Hence it became impossible
to determine how long the source was in the soft state.

\textit{(\romannumeral 5) Declining Hard State (HS):}
After 2020 February 26 (MJD 58905), Swift/XRT data became available for spectral analysis.
The total accretion rate, the $\dot{m}_d$ and the $\dot{m}_h$ all were low in this period.
On the other hand, the ARR became high, and the shock also moved outward at $\sim250$$r_s$ with increased strength.
All of these show that the source had already transitioned into the declining hard state.

\begin{figure*}
\vskip 1.5cm
  \centering
    \includegraphics[angle=0,width=14cm,keepaspectratio=true]{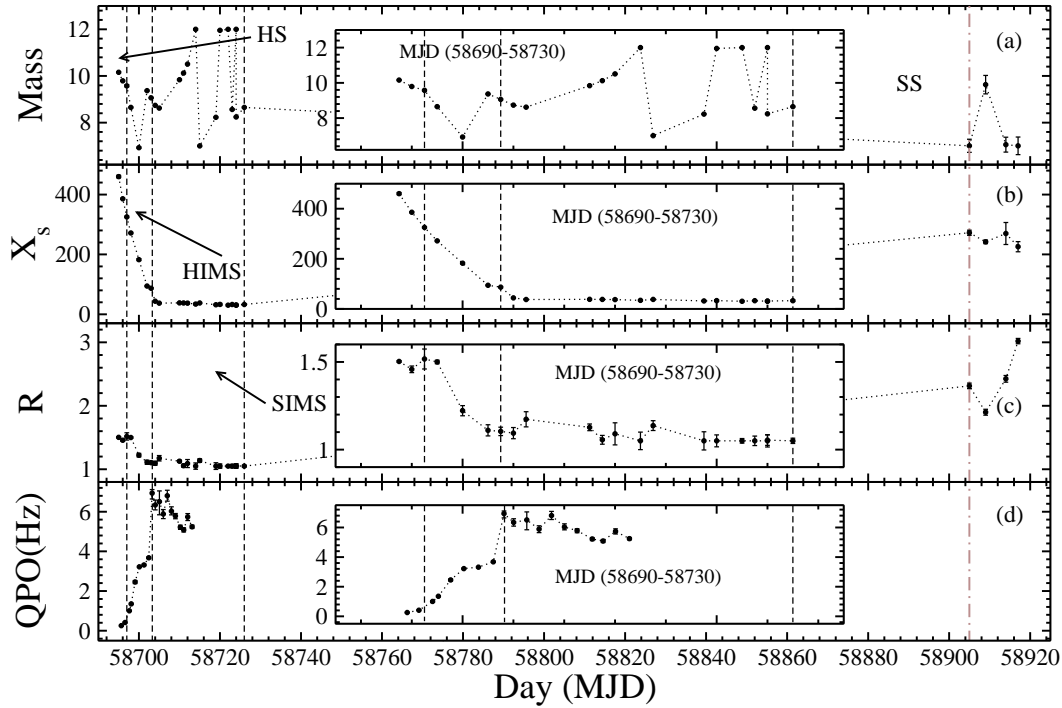}
	\caption{Variations of TCAF model fitted 
	                     (a) mass of the BH ($M_{BH}$) in units of $M_\odot$,
						 (b) the shock location ($X_s$) in units of Schwarzschild radius ($r_s$),
	                     (c) the shock compression ratio ($R$),
					 and (d) the evolution of QPO frequency 
	        are shown with time (MJD). 
                    Vertical dashed (online black) lines mark the transition between spectral states. 
                    The vertical dot-dashed (online pink) line marks the probable start of the declining hard state (see text for details).}
\end{figure*}

\subsubsection{Analysis of Reflection spectra}
We attempted to fit the combined NICER+NuSTAR spectra on MJD 58708 with the absorbed TCAF model.
After adding a {\fontfamily{qcr}\selectfont Gaussian} model to account for the prominent Fe emission line, 
we get a reduced chi-square or $\chi^2_{red}$ of $\sim 1.79$ (Fig. 5a). However, a broad dip around $\sim 10$ keV was 
observed in the residuals, and after adding a {\fontfamily{qcr}\selectfont gabs} model with a line energy 
of $9.71\pm0.23$ keV, we get a $\chi ^2_{red}$ of $\sim 0.91$ (Fig. 5b). This dip could be a sign of 
relativistic disk reflection, as such feature has been found for an earlier NuSTAR observation of this
outburst (Draghis et al. 2020, Wang et al. 2021). To check for this, we used the physical reflection model 
{\fontfamily{qcr}\selectfont relxill} (Garc\'{i}a et al. 2014; Dauser et al. 2014) to further 
analyse the spectrum. We used the {\fontfamily{qcr}\selectfont diskbb} model along with 
{\fontfamily{qcr}\selectfont relxill} to model the thermal emission and the 
{\fontfamily{qcr}\selectfont smedge} model for the low energy instrumental edges. We fixed 
the inner radius to the ISCO value ($r_{in} = 1.0~r_{ISCO}$) and the outer radius ($r_{out}$) 
to $990~GM/c^2$. Although we got a good fit with a $\chi^2_{red}$ $\sim 1.19$, prominent 
residuals remain in the 1-2 keV energy range (Fig. 5c). As signs of reflection are observed 
in the higher energy parts of the spectrum, we ignore the energy range 1-2 keV and fit the 2-79 keV 
spectrum with {\fontfamily{qcr}\selectfont constant*TBabs(diskbb+relxill)} model. This resulted 
in an improved fit with a $\chi^2_{red}$ $\sim 0.83$ which is shown in Fig. 5d. This shows 
that the dip around $\sim 10$ keV is indeed a signature of disk reflection. We obtain a spin parameter
of $a=0.978^{+0.005}_{-0.006}$ and a disk inclination of $67.1^{\circ}\pm0.2^{\circ}$ from our spectral 
fit using the {\fontfamily{qcr}\selectfont relxill} model.

\subsubsection{Estimation of BH mass from spectral analysis}
Mass of the BH ($M_{BH}$) is an important parameter for spectral fitting with TCAF. Mass of the BH 
in EXO 1846-031 was previously determined to be $3.24\pm0.2 ~M_\odot$ based on the presence of 
3:2 ratio HFQPOs (Strohmayer \& Nicer Observatory Science Working Group 2020) and $9\pm5 ~M_\odot$ 
based on spectral modelling (Draghis et al. 2020). Due to the large disparity in the determined mass values,
we tried to estimate it from our spectral modelling. 
Mass of the BH is one of the spectral fitting parameters of the TCAF model, and each spectral fit
gives a best fitted value of the $M_{BH}$ parameter.
We keep the mass parameter free during the analysis with TCAF.
From our spectral fits, we find the best fitted $M_{BH}$ values to vary between $7.1-12.6 ~M_\odot$.
However, mass of a BH in a BHXRB system is not supposed to change significantly during the 
course of an outburst, and the spread in the mass values do not show the variation of the actual BH 
mass. One of the possible reasons for this variation could be that any errors in the measurement of 
the emitted spectra as well as the errors in the fit parameters contribute to proportionally larger 
errors in determination of $M_{BH}$ (Bhattacharjee et al. 2017). Moreover, in our spectral analysis 
we fitted the spectra of different energy bands obtained from multiple instruments of different 
effective areas, which may also contribute to the measurement errors in the mass of the BH. To reduce 
such errors, we perform a global fit using all spectra in different epochs. We use the model 
{\fontfamily{qcr}\selectfont constant*TBabs*smedge(TCAF+Gaussian)} and keep the mass parameter linked 
for all spectra. The parameter space of the fit is very large, and the complexity of the parameter 
space, along with the limited quality of data makes it challenging to explore the parameter space. The fit 
is prone to converging to a local $\chi^2$ minima rather than converging to the global best-fit solution. 
Hence to reduce the complexity of the parameters space, we keep the flow parameters, i.e. the accretion rates, 
the shock location and the compression ratio fixed at their best fitted values from the individual fits. Moreover,
the {\fontfamily{qcr}\selectfont smedge} component is only needed to fit the NICER/XTI spectra, 
and the {\fontfamily{qcr}\selectfont Gaussian} component is required only for the NICER spectra in the rising phase. 
Hence for other spectra, the parameters of the {\fontfamily{qcr}\selectfont smedge} or the {\fontfamily{qcr}\selectfont Gaussian 
components are fixed at zero or the minimum value.} The joint fit is shown in Fig. 8. From the global fit, 
we obtain a mass value of the source as $10.4^{+0.1}_{-0.2}~M_\odot$.

\begin{figure}[h]
\vskip 0.2cm 
  \centering
    \includegraphics[angle=-90,width=9cm,keepaspectratio=true]{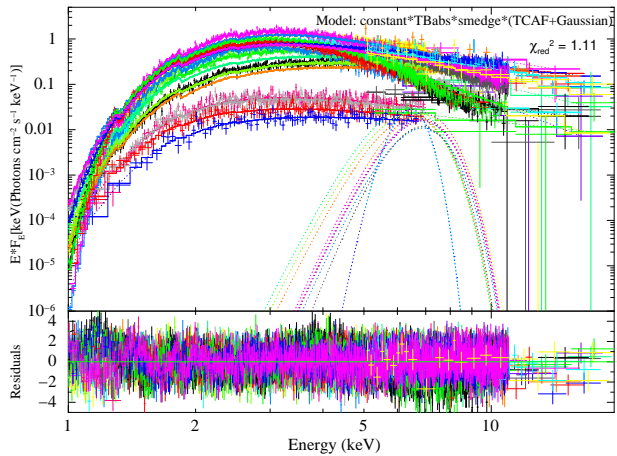}
	\caption{Joint fit of all spectra in different epochs using NICER, Swift and MAXI data.}
\end{figure}

\section{Discussions and Concluding Remarks}
EXO 1846-031 is a galactic black hole candidate that went into an outburst in July 2019 after remaining almost 34 years 
in quiescence after its discovery in 1985. We study the evolution of the temporal and spectral properties of the 
source during the outburst using observational data from Swift/XRT, Swift/BAT, NICER/XTI, MAXI/GSC and NuSTAR/FPM 
satellites/instruments. For the spectral analysis we use the physical TCAF model and fit NICER (1--10 keV), combined
NICER+GSC (1--20 keV), XRT+GSC (1--20 keV) and NICER+NuSTAR (1--79 keV) spectra for 24 epochs spread over the outburst duration.
From our spectral fits, we obtain flow parameters of the system such as the  Keplerian disk accretion rate ($\dot{m}_d$), 
the sub-Keplerian halo accretion rate ($\dot{m}_h$), the shock location ($X_s$), and the shock compression ratio (R). 
As these flow parameters evolve during the outburst, we gain insight into how the accretion flow of matter changes
and produces different kinds of spectral and temporal variability. We also estimate the mass of the black hole from our spectral analysis.

Generally, transient black hole outbursts show two types of outburst profiles, fast rise exponential decay (FRED) or 
slow rise slow decay (SRSD) (Debnath et al. 2010). However, in the case of some BHXRBs, the X-ray flux does not decay 
into quiescence after the first outburst peak. Rather, they show one or more peaks after the main outburst peak
before eventually going into the quiescence phase. In literature, such phenomena are known as ``reflares'', ``rebrightenings''
or ``mini-outbursts'' (e.g. GRO J0422+32, MAXI J1659-152, GRS 1739-278, MAXI J1910-057; 
Chen et al. 1997, Homan et al. 2013, Yan \& Yu 2017, Nath et al. 2023). For the 2019 outburst of EXO 1846-031, we can see from 
Fig. 1 that both the soft and hard fluxes increase rapidly at the start of the outburst. While the hard flux decayed 
slowly after attaining its maximum, the soft flux, though it started to decrease initially, began to increase 
again and attained a maximum comparable with the first peak. This outburst can be classified as a multipeak outburst 
according to the re-brightening classification scheme of Zhang et al. (2019). During the following long declining phase,
two dimmer peaks are also observed. We fitted the 2-10 keV MAXI/GSC outburst profile with multiple FRED models, and from this fit we estimated 
that the total integrated flux released in the outburst is $39.70^{+3.29}_{-3.05}$ Crab day. The contribution of individual peaks calculated from
the individual FRED profiles are 16\%, 78\%, 4\% and 2\% respectively for the first, second, third and fourth peaks. Here we observe that
although the peak fluxes are roughly same, five times more energy is released during the second peak than the first peak.

According to the disk instability model (DIM: see Hameury 2020 for a review), accreting matter from the companion star
accumulates in the disk during quiescence. As more and more matter gets accumulated, it is heated up and gets ionized. 
This ionised matter causes a thermal-viscous instability that causes drastic changes in the viscous and thermodynamic properties 
of the disk, and an outburst is triggered. Although the general quiescence-to-outburst cycle of BHXRBs can be broadly explained with the DIM,
it cannot explain the re-brightenings/mini-outbursts phenomena very well. Although several modifications of DIM have been proposed 
that attempt to explain the reflares  (e.g., Kuulkers et al. 1994; Hameury et al. 2000; Zhang et al. 2019), none of them are well verified. 
Hence we investigate the rebrightening phenomena of EXO 1846-031 with the TCAF picture.

In the TCAF paradigm, viscosity plays a crucial role for triggering an outburst (Chakrabarti et al. 2005, Debnath et al. 2015a, 
Jana et al. 2016 and references therein). For low-mass BHXRBs, the matter supplied by the companion through Roche lobe overflow 
cannot proceed too far inside due to a lack of sufficient viscosity. Hence, rather than getting accumulated in the disk, accreting 
matter starts to pile up at a location called the pile-up radius ($X_p$; Chakrabarti et al. 2019) away from the BH. 
When some instability increases the viscosity at $X_p$,
an increased outward redistribution of angular momentum causes the piled up matter to flow rapidly inward onto the BH, 
triggering an outburst. As $X_p$ is a function of available viscosity, it could vary from one outburst to another.  
It has been shown that this general idea of a variable pile-up location can explain the multiple and anomalous outbursts of low-mass BHXRBs. 
It is observed that anomalous natures of 2003 and 2004 outbursts of the recurring BHXRB H 1743-322 can be explained to be the result of a 
large accumulation of matter at $X_p$ in the preceding long accumulation period ($\sim$ 26 years) that was not fully cleared during the 2003 outburst 
and was partially released during the 2004 outburst (Chakrabarti et al. 2019). For the case of the source 4U 1630-47, multiple different 
types of outbursts ('normal',`mega', and `super' outbursts) can be explained to be the result of a cyclic variation of the pile-up radius 
(Chatterjee et al. 2022). Similar anomalous behavior of the 2013, 2014-15 and 2017–18, 2018–19, 2019–20 outbursts of the source GX 339-4 
has also been explained with the idea of the pile-up radius (Bhowmick et al. 2021). 
Evidently, total matter accumulation at $X_p$ in quiescence and/or the position of $X_p$ varies between different outbursts and different sources. 
Total amount of matter accumulated at $X_p$ depends on the steady matter supply from the companion superposed by anomalous variation of the rate 
of matter supply due to the intrinsic activity of the companion (Chakrabarti et al. 2019). This explains why it took 26 years of accumulation 
before an energetic outburst for H 1743-322 whereas the sources like 4U 1630-47 and GX 339-4 has shown multiple energetic outbursts within similar time period.

From this concept of pile-up radius and the variation of TCAF parameters, the following picture of accretion
dynamics during the 2019 outburst of EXO 1846-031 emerges. Similar to the case of H 1743-322, the long period ($\sim$ 34 years)
of inactivity of EXO 1846-031 indicates a large amount of matter is accumulated at the $X_p$ before the outburst. This accumulated 
matter starts to heat up and gives rise to a convective instability which increases the viscosity due to the 
resulting turbulence. As the viscosity at $X_p$ increases above the critical value, the outburst is triggered and matter begins to rush inward. 
We can see from Fig. 6 that the halo rate is high compared to the disk rate initially.
As the sub-Keplerian matter has low viscosity, it falls freely towards the BH, whereas the Keplerian matter 
has large viscosity and it moves inward slower in viscous timescale. The sub-Keplerian matter reaches 
the BH faster than the Keplerian matter, and the halo rate attains peak value before the disk rate. 
From Fig. 7, we see that the shock is far away in this initial phase. As there is no Keplerian matter to cool 
the faster-moving sub-Keplerian matter, it forms a large CENBOL, and this CENBOL inverse Comptonizes most of
the soft thermal photons and produces a large number of hard photons. Hence we can see from Fig. 1 that the 
high energy fluxes dominate the low energy fluxes making the HRs high and the source goes into the rising 
hard state. After MJD 58697, the Keplerian matter begins to cool down the sub-Keplerian matter as it 
gradually moves towards the BH. The disk rate starts to increase and the halo rate decreases. The CENBOL
shrinks in size, and the shock, which is the outer boundary of CENBOL, moves closer to the BH and decreases 
in strength. Hence the inverse-Comptonization is reduced, the hard flux begins to decrease, the thermal flux
increases, the HRs decrease, and the source goes into the hard intermediate state. 
As the accumulated matter depletes, viscosity at $X_p$ drops before the release of all accumulated matter. 
The supply of accreting matter gradually decreases, both the disk rate and halo rate decrease, and the CENBOL shrinks 
further and the shock moves very closer to the BH. Both the soft and the hard flux decrease, the HRs are decreased to a very low level
and the source goes into a soft-intermediate state. As the sub-Keplerian matter moves faster, all of it 
gets accreted quickly as the source enters the SIMS, which could also be interpreted as the declining state 
of the first failed (as the soft state is not reached) outburst.
In all of these three states, we find the presence of low-frequency 
quasi-periodic oscillations (LFQPO). In the TCAF picture, LFQPOs are produced due to the oscillation of
the shock, i.e. the outer boundary of the CENBOL. The centroid frequency of the LFQPO ($\nu_{QPO}$) is roughly inversely
proportional to the location of the shock ($r_s$) ($\nu_{QPO} \sim 1/r_s(r_s-1)^{1/2}$: Chakrabarti et al. 2008, Debnath et al. 2010).
As we can see from Fig. 7, as the shock moves closer to the BH in the HS and HIMS, the centroid frequency of the QPO
increases. As the shock becomes almost steady in the SIMS, the QPO frequency also becomes steady.
After some days in the SIMS ($\sim$ MJD 58715), the value of the compression ratio becomes close to one
and the halo rate becomes close to zero. This signifies that the post-shock and pre-shock matter density is equal, 
which means that the shock has become very weak or disappeared and it has moved very close to the black hole.
As the shock disappears, the sub-Keplerian and Keplerian components of the accretion flow becomes essentially the same.
This very weak shock is unable to produce any QPOs, hence the QPO disappear gradually at the later stage of the SIMS.
 
As the companion supplies matter continuously, new matter is accumulated at $X_p$ though it is not released due to
a lack of viscosity. After MJD 58726, viscosity at the $X_p$ increases again due to the onset of some instability, and the 
remaining matter is released triggering the reflare event. The soft fluxes began to increase again while the hard fluxes 
remained low, which shows that there is an increase in the thermal emission without much of it being inverse-Comptonized.
Although some NICER and XRT spectra are available in this phase, we failed to fit these spectra with the TCAF model. 
These indicate that the two component configuration of the accretion flow is no longer maintained in this period.
Majority of the sub-Keplerian matter was accreted during the first peak, and this new accretion consists largly of high viscous 
Keplerian matter. This Keplerian matter interacts with the remaining small amounts of sub-Keplerian matter and cools it down
and only the Keplerian disk accretion occurs in this state. 
From Fig. 1, we can see that after attaining the second maximum, the soft flux decreases almost linearly 
instead of an exponential decline, which is another indication that only the comparatively slow moving Keplerian
matter is responsible for this reflare. After $\sim$ MJD 58800, the source became Sun-constrained 
and there is no available data for spectral analysis in the period between MJD 58808 and MJD 58904.
Hence the exact date when the source came out of the SS cannot be determined. 
After MJD 58905, spectral analysis shows that the shock has moved outward at $\sim 250 r_s$ with an 
increased ARR. This indicates the source has already moved into the declining hard state.

Wang et al. (2021) has detected a highly ionised disk wind with a velocity up to 0.06c 
in this source. Such disk winds could be produced by radiation pressure (Proga \& Kallman 2002, 
Higginbottom \& Proga 2015). This radiation could irradiate the remaining matter at $X_p$, and 
this can be a possible reason for the onset of the instability that increases the viscosity at 
$X_p$ triggering the reflare. Similar phenomena was observed during the 2002-03 and 2004-05 
outbursts of the source GX 339-4 where a second outburst peak was observed due to an instability 
that was triggered in the accretion disk due to the irradiation from the central X-ray source after 
the primary peak (Aneesha et al. 2019).

The time taken by the high viscous matter to reach the BH from the pile-up radius is termed 
as the viscous timescale (Chakrabarti et al. 2019). Due to its low viscosity, the sub-Keplerian matter
moves toward the BH in freefall timescale, whereas the Keplerian matter takes more time to reach the BH due 
to its higher viscosity. Due to this reason, it is generally observed that the halo accretion rate attains 
its peak before the disk rate, and the time difference between disk and halo peaks gives us an idea to infer 
viscous timescale of the source (Debnath et al. 2015b, Jana et al. 2016, 2020b). From Fig. 1, we can see 
that 15-50 keV Swift/BAT hard flux attains a peak on MJD 58796, and the 2-4 keV MAXI/GSC soft flux attains a peak 
$\sim 8$ days later on MJD 58705. A similar delay between the peaks of halo and disk rates is also found (see Fig. 6). 
Hence we estimate the viscous timescale of this source to be $\sim 8$ days. This large viscous timescale indicates
that $X_p$ is far away from the BH and size of the accretion disk is large.

The mass of the BH in this source has not yet been measured dynamically, so we try to estimate the mass
from our spectral fits. The spectra emitted from the accretion processes around a BH is highly dependent 
on its mass as various features of the accretion dynamics such as the size of the CENBOL and the electron 
number density inside it, soft radiation intensity from the Keplerian disk, etc. depend on the mass. We allow
the $M_{BH}$ parameter to vary freely during our spectral analysis so that we get a best fitted value of the
parameter from each spectral fit. We find the best fitted values of the parameter to vary in the range 
$7.1-12.6 ~M_\odot$. This spread in the mass value may be a consequence of systematic errors 
due to the use of data from multiple instruments with different energy range and effective areas.
To reduce such errors, we jointly fit all the spectra of different epochs and estimate the most probable 
mass of the source to be $10.4^{+0.1}_{-0.2}~M_\odot$.

\section{Summary and Conclusions}
We study the spectral and temporal properties of the source EXO 1846-031 during its 2019 outburst after almost 34
years in quiescence. We use MAXI/GSC and Swift/BAT daily lightcurve data to study the evolution of the X-ray
fluxes and hardness ratios during the outburst. We use the FRED profile to fit the outburst flux profile and
estimate the contribution of each flux peak in the total amount of flux released during the outburst.
We use data from multiple instruments (Swift/XRT, MAXI/GSC, NICER/XTI, NuSTAR/FPM) for a broadband spectral study over a period of 222 days. 
We perform our spectral study using physical TCAF model. Based on our spectral analysis, the following conclusions can be drawn:

\begin{itemize}

\item After 34 years in quiescence, EXO 1846-031 showed an outburst in 2019 that can be classified as a multipeak outburst.

\item Before the start of the outburst, a large amount of matter was accumulated at the pile up radius $X_p$ (located far away 
from the BH) and all the matter was not accreted during the first outburst peak.

\item The broad absorption feature around $\sim 9$ keV indicates the presence of a fast moving highly ionized disk wind in the rising SIMS.

\item As the high X-ray flux irradiates the remaining matter at $X_p$, the viscosity increases and starts a fresh accretion of 
matter triggering the reflare. 

\item This increased supply of high viscous Keplerian matter in the reflaring event cools down and washes off the sub-Keplerian matter, 
and only Keplerian disk accretion happens in the HSS.

\item Although the source showed two brighter and two dimmer peaks during the outburst, $\sim 78\%$ of total energy has been released in the second 
flaring event.

\item From spectral fitting with TCAF, probable mass of the source was estimated to be $10.4^{+0.1}_{-0.2}~M_\odot$.

\item From the disk and halo peak difference in the rising phase of the outburst, we estimated viscous time scale of the source to be $\sim 8$ days.

\end{itemize}

\section*{Acknowledgements}

This work made use of Swift/XRT, Swift/BAT, NICER/XTI, and NuSTAR/FPM data supplied by the UK Swift Science Data Centre at the University of Leicester, 
and MAXI/GSC data were provided by RIKEN, JAXA, and the MAXI team. S.K.N. acknowledges support from the SVMCM fellowship, the government of West Bengal.
S.K.N. and D.D. acknowledge support from the ISRO-sponsored RESPOND project (ISRO/RES/2/418/17-18) fund. D.D. and K.C. acknowledge visiting research grants 
of National Tsing Hua University, Taiwan (NSTC 111-2811-M-007-066). R.B. acknowledges support from the CSIR-UGC fellowship (June-2018, 527223).
H.-K. C. is supported by NSTC of Taiwan under grant 111-2112-M-007-019.

\begin{table}
\small 
 \addtolength{\tabcolsep}{-3.0pt}
 \centering
\caption{Log of Swift, NICER, MAXI and NuSTAR observations of the transient BHC EXO~1846-031.\label{tab1}}
\begin{tabular}{ccccccccc}
\hline\hline
ID   &  Obs. ID    &   Satellite/Instrument  &   MJD    &   Date of Obs.  &  XRT & NICER & GSC & NuSTAR\\
     &             &                         &          &  (YYYY-MM-DD)     &   Exposure(s)  & Exposure(s) & Exposure(s) & Exposure(s)\\
(1)  &    (2)      &          (3)            &   (4)    &   (5)           &   (6) & (7) & (8) & (9) \\
\hline\hline
X1   & 00011500002 &  XRT+GSC  & 58697 &  2019-08-02  & 1138.070 & --- &  2840.156  & ---\\ 
X2   & 00011500003 &  XRT+GSC  & 58700 &  2019-08-05  & 1220.714 & --- &  3508.613  & ---\\ 
X3   & 00011500006 &  XRT+GSC  & 58703 &  2019-08-08  &  862.014 & --- &   121.049  & ---\\ 
X4   & 00011500008 &  XRT+GSC  & 58705 &  2019-08-10  & 1339.248 & --- &   967.348  & ---\\
X5   & 00011500016 &  XRT+GSC  & 58715 &  2019-08-20  & 1329.173 & --- &   420.399  & ---\\
X6   & 00011500017 &  XRT+GSC  & 58719 &  2019-08-23  &  884.042 & --- &   524.238  & ---\\
X7   & 00088981001 &  XRT+GSC  & 58723 &  2019-08-27  &  839.121 & --- &  1970.271  & ---\\
X8   & 00011500030 &  XRT+GSC  & 58905 &  2020-02-26  &  909.452 & --- &  3042.624  & ---\\
X9   & 00011500031 &  XRT+GSC  & 58909 &  2020-03-01  &  884.481 & --- &  2090.291  & ---\\
X10  & 00011500032 &  XRT+GSC  & 58914 &  2020-03-05  &  919.454 & --- &  3405.380  & ---\\
X11  & 00011500033 &  XRT+GSC  & 58917 &  2020-03-09  &  939.214 & --- &  3663.832  & ---\\
\hline
NI1  & 2200760101  &    NICER  & 58695 &  2019-07-31  & --- & 5658.000 &  ---  & --- \\
NI2  & 2200760102  &    NICER  & 58696 &  2019-08-01  & --- & 1165.000 &  ---  & --- \\
NI3  & 2200760104  &    NICER  & 58698 &  2019-08-03  & --- & 1488.000 &  ---  & --- \\
NI4  & 2200760108  & NICER+GSC & 58702 &  2019-08-07  & --- &  927.000 & 3508.613 & --- \\
NI5  & 2200760110  &    NICER  & 58704 &  2019-08-09  & --- & 4629.000 &  ---  & --- \\
NI6  & 2200760116  & NICER+GSC & 58710 &  2019-08-15  & --- & 4703.000 & 1440.399 & --- \\
NI7  & 2200760117  &    NICER  & 58711 &  2019-08-16  & --- & 8439.000 &  ---  & --- \\
NI8  & 2200760118  &    NICER  & 58712 &  2019-08-17  & --- & 4875.000 &  ---  & --- \\
NI9  & 2200760120  &    NICER  & 58714 &  2019-08-19  & --- & 3894.000 &  ---  & --- \\
NI10 & 2200760126  &    NICER  & 58720 &  2019-08-25  & --- & 1129.000 &  ---  & --- \\
NI11 & 2200760128  &    NICER  & 58722 &  2019-08-27  & --- & 1501.000 &  ---  & --- \\
NI12 & 2200760130  &    NICER  & 58724 &  2019-08-29  & --- &   79.000 &  ---  & --- \\
NI13 & 2200760132  &    NICER  & 58726 &  2019-08-31  & --- &  384.000 &  ---  & --- \\
\hline
NU1  & 2200760114 & NICER+NuSTAR & 58708 & 2019-08-13 & --- & 7154.000 & --- & 17672.0 \\ 
\hline\hline
\end{tabular}
\end{table}

\begin{table}
 \footnotesize
 \addtolength{\tabcolsep}{-3.0pt}
 \centering
 \caption{Best fitted spectral model parameters.\label{tab2}}
                \begin{tabular}{ccccccccccc}
\hline\hline
     Day    &  $\dot{m}_d$$^a$  &  $\dot{m}_h$$^a$  & $X_s$$^a$ & $R$$^a$ & $M_{BH}$$^a$    &   LE$^b$   &   LW$^b$   &   LN$^b$   & $N_H$$^c$ &  $\chi^2/dof$$^d$  \\
	(MJD)   & ($\dot{M}_{Edd}$) & ($\dot{M}_{Edd}$) &  $(r_s)$  &         & ($M_\odot$) &  (keV)  &  (keV)  &  (photons cm$^{-2}$ s$^{-1}$)    &  &                   \\
     (1)    &        (2)        &         (3)       &    (4)    &   (5)   &   (6)         &         (7)             &      (8)     &   (9)   &   (10)    &     (11)     \\
\hline\hline
58695 & $ 0.301^{\pm0.013} $ & $ 0.947^{\pm0.015} $ & $ 460.000^{\pm4.755}$ & $ 1.503^{\pm0.005} $ & $ 10.158^{\pm0.030} $ & $ 6.42^{\pm0.06} $ & $ 0.80^{\pm0.17} $ & $ 0.006^{\pm0.001} $ & $ 5.916^{\pm0.185} $ & $  934.54/976 $ \\ 
58696 & $ 0.305^{\pm0.006} $ & $ 1.002^{\pm0.009} $ & $ 385.887^{\pm1.871}$ & $ 1.458^{\pm0.019} $ & $  9.791^{\pm0.063} $ & $ 6.30^{\pm0.12} $ & $ 0.80^{\pm0.38} $ & $ 0.007^{\pm0.003} $ & $ 5.927^{\pm0.318} $ & $  847.79/880 $ \\ 
58697 & $ 0.342^{\pm0.014} $ & $ 0.711^{\pm0.021} $ & $ 325.417^{\pm2.325}$ & $ 1.517^{\pm0.057} $ & $  9.576^{\pm0.117} $ & $    ---         $ & $     ---        $ & $     ---          $ & $ 6.791^{\pm1.018} $ & $  650.58/591 $ \\ 
58698 & $ 0.425^{\pm0.010} $ & $ 0.527^{\pm0.006} $ & $ 271.496^{\pm1.901}$ & $ 1.500^{\pm0.010} $ & $ 10.301^{\pm0.042} $ & $ 6.28^{\pm0.08} $ & $ 0.80^{\pm0.22} $ & $ 0.011^{\pm0.003} $ & $ 5.814^{\pm0.203} $ & $  930.53/917 $ \\ 
58700 & $ 0.559^{\pm0.010} $ & $ 0.322^{\pm0.011} $ & $ 182.503^{\pm3.446}$ & $ 1.222^{\pm0.029} $ & $  6.926^{\pm0.061} $ & $    ---         $ & $     ---        $ & $     ---          $ & $ 6.902^{\pm1.084} $ & $  742.81/635 $ \\ 
58702 & $ 0.761^{\pm0.014} $ & $ 0.235^{\pm0.012} $ & $  94.264^{\pm1.361}$ & $ 1.110^{\pm0.032} $ & $  9.369^{\pm0.054} $ & $ 6.40^{\pm0.08} $ & $ 0.41^{\pm0.10} $ & $ 0.005^{\pm0.001} $ & $ 6.578^{\pm0.310} $ & $  944.46/905 $ \\ 
58703 & $ 0.772^{\pm0.011} $ & $ 0.236^{\pm0.012} $ & $  86.931^{\pm2.054}$ & $ 1.104^{\pm0.024} $ & $  9.063^{\pm0.105} $ & $    ---         $ & $     ---        $ & $     ---          $ & $ 7.998^{\pm1.745} $ & $  664.96/584 $ \\ 
58704 & $ 0.782^{\pm0.007} $ & $ 0.120^{\pm0.016} $ & $  43.776^{\pm1.624}$ & $ 1.094^{\pm0.032} $ & $  8.737^{\pm0.091} $ & $ 6.80^{\pm0.40} $ & $ 0.80^{\pm0.49} $ & $ 0.004^{\pm0.002} $ & $ 8.619^{\pm0.336} $ & $  872.50/976 $ \\ 
58705 & $ 0.611^{\pm0.013} $ & $ 0.109^{\pm0.011} $ & $  37.341^{\pm1.382}$ & $ 1.173^{\pm0.043} $ & $  8.621^{\pm0.070} $ & $    ---         $ & $     ---        $ & $     ---          $ & $ 9.953^{\pm1.362} $ & $  717.50/616 $ \\ 
58710 & $ 0.391^{\pm0.030} $ & $ 0.109^{\pm0.010} $ & $  38.062^{\pm1.922}$ & $ 1.127^{\pm0.019} $ & $  9.839^{\pm0.032} $ & $ 6.80^{\pm0.16} $ & $ 0.80^{\pm0.19} $ & $ 0.007^{\pm0.001} $ & $ 7.447^{\pm0.401} $ & $  917.96/974 $ \\ 
58711 & $ 0.430^{\pm0.007} $ & $ 0.111^{\pm0.007} $ & $  37.910^{\pm2.017}$ & $ 1.056^{\pm0.025} $ & $ 10.123^{\pm0.031} $ & $ 6.80^{\pm0.16} $ & $ 0.80^{\pm0.19} $ & $ 0.005^{\pm0.001} $ & $ 8.023^{\pm0.181} $ & $  886.27/984 $ \\ 
58712 & $ 0.391^{\pm0.011} $ & $ 0.109^{\pm0.013} $ & $  37.218^{\pm3.115}$ & $ 1.090^{\pm0.063} $ & $ 10.507^{\pm0.036} $ & $ 6.80^{\pm0.18} $ & $ 0.80^{\pm0.22} $ & $ 0.006^{\pm0.001} $ & $ 7.623^{\pm0.272} $ & $  881.32/966 $ \\ 
58714 & $ 0.400^{\pm0.025} $ & $ 0.082^{\pm0.012} $ & $  34.243^{\pm3.281}$ & $ 1.050^{\pm0.050} $ & $ 11.903^{\pm0.103} $ & $ 6.80^{\pm0.29} $ & $ 0.73^{\pm0.18} $ & $ 0.003^{\pm0.001} $ & $ 9.133^{\pm0.303} $ & $  859.55/942 $ \\ 
58715 & $ 0.435^{\pm0.007} $ & $ 0.065^{\pm0.023} $ & $  37.930^{\pm1.262}$ & $ 1.137^{\pm0.028} $ & $  7.830^{\pm0.075} $ & $    ---         $ & $     ---        $ & $     ---          $ & $ 9.564^{\pm2.042} $ & $  636.04/575 $ \\ 
58719 & $ 0.352^{\pm0.015} $ & $ 0.050^{\pm0.019} $ & $  31.855^{\pm2.143}$ & $ 1.050^{\pm0.052} $ & $  8.229^{\pm0.134} $ & $    ---         $ & $     ---        $ & $     ---          $ & $ 8.643^{\pm1.209} $ & $  542.37/500 $ \\ 
58720 & $ 0.385^{\pm0.005} $ & $ 0.052^{\pm0.030} $ & $  32.958^{\pm4.007}$ & $ 1.050^{\pm0.034} $ & $ 11.956^{\pm0.179} $ & $    ---         $ & $     ---        $ & $     ---          $ & $10.937^{\pm0.471} $ & $  782.03/757 $ \\ 
58722 & $ 0.351^{\pm0.021} $ & $ 0.050^{\pm0.014} $ & $  30.562^{\pm2.511}$ & $ 1.050^{\pm0.013} $ & $ 12.570^{\pm0.068} $ & $    ---         $ & $     ---        $ & $     ---          $ & $ 7.788^{\pm0.339} $ & $ 1307.11/745 $ \\ 
58723 & $ 0.374^{\pm0.019} $ & $ 0.052^{\pm0.006} $ & $  32.679^{\pm1.382}$ & $ 1.050^{\pm0.027} $ & $  8.563^{\pm0.186} $ & $    ---         $ & $     ---        $ & $     ---          $ & $ 8.852^{\pm0.758} $ & $  533.80/501 $ \\ 
58724 & $ 0.351^{\pm0.013} $ & $ 0.050^{\pm0.008} $ & $  30.321^{\pm2.627}$ & $ 1.050^{\pm0.035} $ & $ 12.352^{\pm0.414} $ & $    ---         $ & $     ---        $ & $     ---          $ & $ 7.891^{\pm1.121} $ & $  525.94/474 $ \\ 
58726 & $ 0.360^{\pm0.014} $ & $ 0.050^{\pm0.017} $ & $  32.793^{\pm2.941}$ & $ 1.050^{\pm0.015} $ & $  8.616^{\pm0.141} $ & $    ---         $ & $     ---        $ & $     ---          $ & $ 7.838^{\pm0.428} $ & $ 1095.62/639 $ \\ 
58905 & $ 0.058^{\pm0.002} $ & $ 0.229^{\pm0.005} $ & $ 272.479^{\pm8.555}$ & $ 2.314^{\pm0.043} $ & $  7.232^{\pm0.281} $ & $    ---         $ & $     ---        $ & $     ---          $ & $ 5.122^{\pm5.306} $ & $  195.28/176 $ \\ 
58909 & $ 0.075^{\pm0.002} $ & $ 0.198^{\pm0.008} $ & $ 241.968^{\pm6.093}$ & $ 1.899^{\pm0.047} $ & $  9.627^{\pm0.395} $ & $    ---         $ & $     ---        $ & $     ---          $ & $ 5.743^{\pm2.279} $ & $  140.21/159 $ \\ 
58914 & $ 0.037^{\pm0.002} $ & $ 0.245^{\pm0.003} $ & $ 269.825^{\pm36.74}$ & $ 2.425^{\pm0.053} $ & $  7.055^{\pm0.319} $ & $    ---         $ & $     ---        $ & $     ---          $ & $ 5.071^{\pm4.441} $ & $  103.17/124 $ \\ 
58917 & $ 0.020^{\pm0.002} $ & $ 0.300^{\pm0.007} $ & $ 224.856^{\pm16.91}$ & $ 3.019^{\pm0.042} $ & $  7.620^{\pm0.377} $ & $    ---         $ & $     ---        $ & $     ---          $ & $ 5.368^{\pm2.648} $ & $   72.75/82  $ \\ 
\hline\hline
                \end{tabular}
\vskip 0.2cm
\noindent{
        \leftline{$^a$ TCAF model spectral parameters are shown in columns (2)-(6).} 
        \leftline{$^b$ Line energy (LE), line width (LW) and line normalization (LN) of the Gaussian component are shown in columns (7)-(9).}
        \leftline{$^c$ Equivalent hydrogen column density ($N_H$) values are shown in column (10) in units of $10^{22}$ atoms cm$^{-2}$.}
	\leftline{$^d$ $\chi^2$ values and degrees of freedom (dof) of the TCAF spectral model fits are shown in column (11).}
        \leftline{\textit{Note.} Superscripts on the parameter values represent average error values obtained using the {\fontfamily{qcr}\selectfont err}
                           task in {\fontfamily{qcr}\selectfont XSPEC} with 90\% confidence.}

}

\end{table}

\end{document}